\documentclass[10pt,prl,aps,superscriptaddress,twocolumn,amsmath,amssymb]{revtex4-1}

\usepackage{siunitx}
\usepackage{graphicx}
\usepackage{amsmath}
\usepackage{dcolumn}
\usepackage{lineno}
\usepackage[utf8]{inputenc}

\begin{document}
\widetext
\title{Dumbbell impurities in 2D crystals of repulsive colloidal spheres induce particle-bound dislocations}

\author{Vera Meester}
\affiliation{Soft Matter Physics, Huygens-Kamerlingh Onnes Laboratory, Leiden University, PO Box 9504, 2300 RA Leiden, The Netherlands}
\author{Casper van der Wel}
\affiliation{Soft Matter Physics, Huygens-Kamerlingh Onnes Laboratory, Leiden University, PO Box 9504, 2300 RA Leiden, The Netherlands}
\author{Ruben W. Verweij}
\affiliation{Soft Matter Physics, Huygens-Kamerlingh Onnes Laboratory, Leiden University, PO Box 9504, 2300 RA Leiden, The Netherlands}
\author{Giovanni Biondaro}
\affiliation{Soft Matter Physics, Huygens-Kamerlingh Onnes Laboratory, Leiden University, PO Box 9504, 2300 RA Leiden, The Netherlands}
\author{Daniela J. Kraft}
\email{kraft@physics.leidenuniv.nl}
\affiliation{Soft Matter Physics, Huygens-Kamerlingh Onnes Laboratory, Leiden University, PO Box 9504, 2300 RA Leiden, The Netherlands}

\begin{abstract}
Impurity-induced defects play a crucial role for the properties of crystals, but little is known about impurities with anisotropic shape. Here, we study how colloidal dumbbells distort and interact with a hexagonal crystal of charged colloidal spheres at a fluid interface. We find that subtle changes in the dumbbell length induce a transition from a local distortion to a particle-bound dislocation, and determine how the dumbbell moves inside the repulsive hexagonal lattice. Our results provide new routes towards controlling material properties through particle-bound dislocations.
\end{abstract}
	\maketitle
\small 

Defects can distort the crystalline order both locally through a slight deviation from the lattice positions and over long distances by introducing dislocations or grain boundaries.\cite{Bollman1970} They can be introduced by vacancies or interstitials resulting in point defects\cite{Lechner2008}, by substrate curvature inducing topological defects \cite{Irvine2010, Meng2014, Wales2014, Ershov2013, Kusumaatmaja2013}, or due to the presence of impurities or dopants.\cite{DeVilleneuve2005} These imperfections in the crystalline order can influence the mechanical strength of the crystal\cite{Wood2011}, change the crystal growth direction\cite{Gasser2009}, enhance or deteriorate photonic properties\cite{Braun2006,Lopez2003} and induce melting.\cite{Alsayed2005} Defects are thus intricately linked to and can be used as a tool to adapt the properties of a material.

Doping a crystal with a second component is a straightforward means to control the formation of defects and hence the material properties. To monitor the influence of such impurities on the crystalline order, colloidal particles are a convenient choice because they allow for real-time studies at the particle level. For example, large, spherical impurities in colloidal crystals have been found to inhibit crystal growth by inducing the formation and segregation of grain boundaries.\cite{DeVilleneuve2005, Lavergne2016}  Surprisingly, much less is known about the influence of anisotropically shaped impurities, while molecular impurities are typically not spherically symmetric \cite{Ambrose:1991} and biological systems such as immature virus particles are known to feature anisotropic defects.\cite{Briggs2009} 
Moreover, close-packed crystals of spheres doped with only a low fraction of dumbbells, i.e. particles consisting of two connected spheres, already exhibited unexpected caging of dislocations and glassy like dislocation dynamics. \cite{Gerbode2008, Gerbode2010} This indicates the potential impact that controlled doping with anisotropic impurities can have on the mechanical properties of materials. However, it is completely unexplored how even this simplest type of impurity, the dumbbell, interacts with and distorts a crystal of spheres when it does not exactly substitute two lattice spacings and can exhibit, albeit limited, translational and rotational motion. 

Here, we relieve this steric constraint by using crystals and impurities that electrostatically repel each other over distances  several times their diameter.  This allows us to study in detail how dumbbells that occupy 1, 1.5, and 2 lattice sites dynamically align with and distort the hexagonal order of two-dimensional crystals. We furthermore investigate how the surrounding crystal constrains the dumbbell motion at various lattice spacings. 

Our colloidal systems consists of 2 $\mu$m-sized, negatively charged polymethylmethacrylate (pMMA) spheres and dumbbells attached to a planar fluid interface between an organic phase consisting of 70:30 cyclohexylbromide:cis-decaline and an aqueous phase made up of 85:15 glycerol water (see Figure S1 for our experimental setup). In the absence of dumbbell impurities, the spherical particles (diameter $d$) position on a hexagonal lattice (Figure S2). We image the dynamics of the colloidal particles whose core was fluorescently labelled using rhodamine-aminostyrene \cite{Dullens:2003, Elsesser:2010} using an inverted Nikon Ti-E confocal microscope equipped with a 60x long working distance objective (NA 0.7). We extract their positions using a modified version of the Crocker-Grier algorithm that can track dumbbells with overlapping features.\cite{Crocker1996, trackpy:vs, vanderWel2017} 
The dumbbells spontaneously form during sample preparation and consist of the spheres that also make up the crystal. They are likely held together by aqueous droplets dispersed in the oil phase with a size below the diffraction limit. The interparticle distance $s_{DB}$ is constant for individual dumbbells (Figure S3). However, $s_{DB}$ varies between different dumbbells allowing us to study the defect type and distortion range they induce in the surrounding hexagonal crystal as a function of their length.

\begin{figure}[t!]
	\centering
	\includegraphics[width=0.3\textwidth]{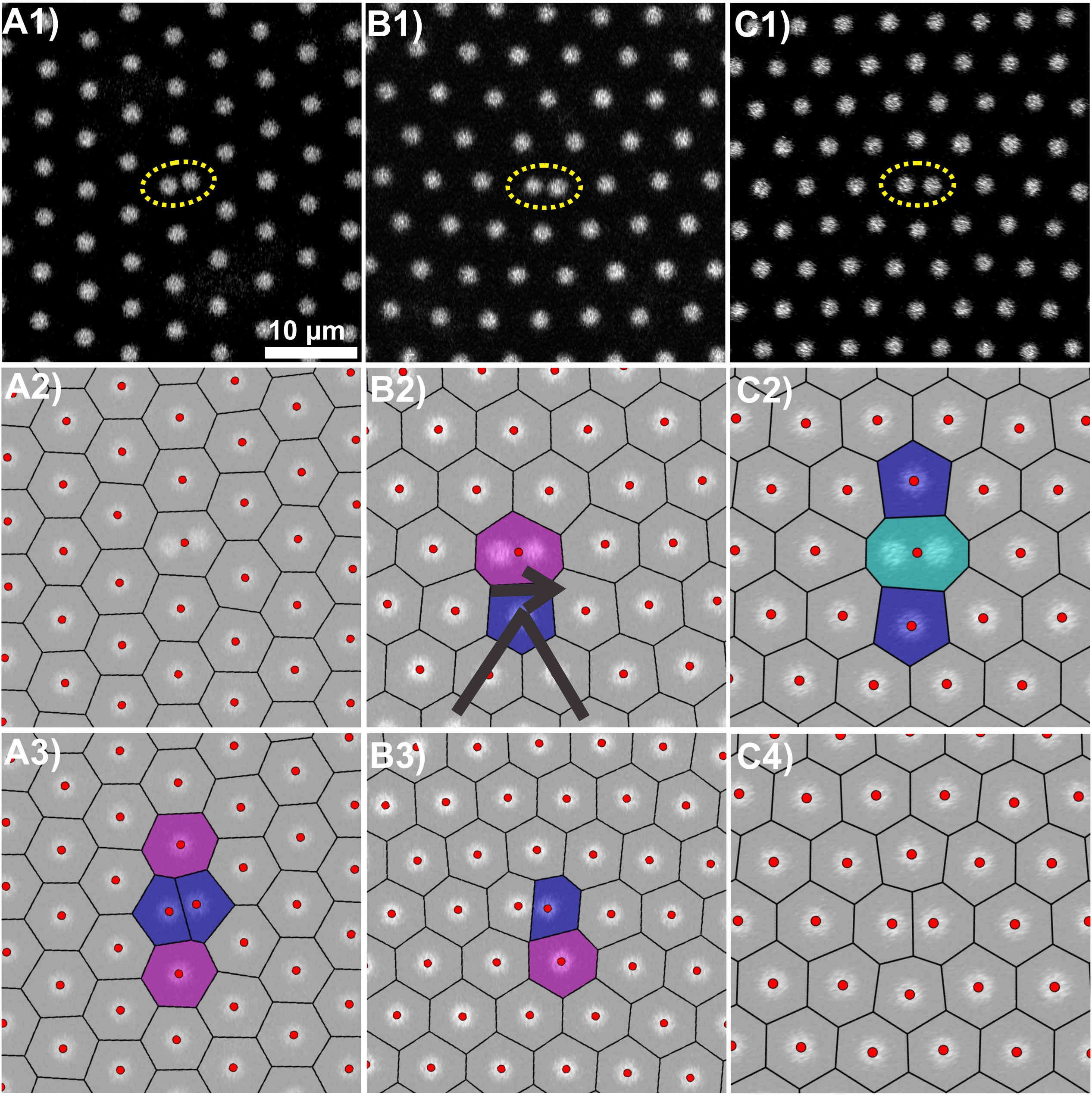}
	\caption{\footnotesize Number of nearest neighbors (NNs) of and defects induced by dumbbell impurities with different lengths $s_{DB}$ in a hexagonal crystal of highly repulsive spheres. 
	Confocal microscopy images of dumbbells with A1) $s_{DB}$ = 1.17$d$ and six NNs, B1) $s_{DB}$ = 1.29$d$ and seven NNs, and C1) $s_{DB}$ = 1.43$d$ and eight NNs, where $d$ is the diameter of the spheres.  Voronoi diagrams of the microscopy images where the dumbbell is considered (A2-C2) as one particle and (A3-C3) as two individual spheres. The Voronoi cells are colored according to the coordination number, $z_{i}$, of the colloids; purple for $z_{i}$=5, grey for $z_{i}$=6, magenta for $z_{i}$=7 and turquoise for $z_{i}$=8. A short dumbbell with six NNs substitutes one lattice position in the hexagonal crystal without the formation of dislocations (A4). At seven NNs a five-and-seven disclination pair forms, resulting in the insertion of a semi-infinite row in the hexagonal crystal (B2 and B4). The arrow indicates the Burgers vector. At eight NNs the dumbbell occupies two lattice sites in the hexagonal crystal (C4).}
	\label{fig:Voronoi}
\end{figure}

The first striking observation is that the number of nearest neighbours (NNs) and hence the type of induced defect varies strongly with a subtle change in the dumbbell length, $s_{DB}$. By drawing the Voronoi diagrams, we can identify the number of NNs and type of defects induced by the presence of the dumbbells. Analysis of 250 dumbbells showed that dumbbells with $s_{DB}$ = (1.03 $\pm$ 0.19)$d$ were surrounded by six NNs, dumbbells with $s_{DB}$ = (1.27 $\pm$ 0.14)$d$ by seven spheres and at $s_{DB}$ = (1.42 $\pm$ 0.17)$d$ by eight, see Fig. \ref{fig:Voronoi}A1-2, B1-2, and C1-2, respectively, and Figure S3. In this dumbbell-centric view, dumbbells with a coordination number of  $z_{i}$=6 substitute one lattice site in the hexagonal crystal without long-ranged distortion of the orientational and translational order (Fig. \ref{fig:Voronoi}A2). 
For dumbbells of intermediate length and $z_{i}$=7, the Voronoi diagrams expose a five-and-seven coordinated disclination pair (Fig. \ref{fig:Voronoi}B2). Isolated dislocations distort the translational, but not the rotational order of hexagonal crystals by the insertion of a semi-infinite row in the crystal, as illustrated by the black lines in Fig. \ref{fig:Voronoi}B2. The dumbbell's long axis is aligned parallel to the Burgers vector (Fig. S4).
At slightly larger $s_{DB}$, the dumbbell itself hosts a coordination number of $z_{i}$=8 and two of the NNs have $z_{i}$=5, which is a rather unusual dislocation (Fig. \ref{fig:Voronoi}C2).

This unusual dislocation vanishes if we consider the dumbbell as two individual spheres when drawing the Voronoi diagram. Now, the individual spheres of the longest dumbbells substitute two lattice positions in the hexagonal crystal, without long-ranged distortions of the crystal (Figure \ref{fig:Voronoi}C3). In this view, the shortest dumbbells reveal a dislocation pair, with $z_{i}$=5 at the spheres forming the dumbbell and $z_{i}$=7 at two of the NNs (Figure \ref{fig:Voronoi}A3) without a long-ranged distortion of the orientational and translational order. Finally, dumbbells of intermediate lengths exhibit a disclination pair, with $z_{i}$=5 at one of the spheres forming the dumbbell and $z_{i}$=7 at one of the neighboring spheres (Figure \ref{fig:Voronoi}C3). This alternative Voronoi diagram shows that the dislocation is spawned by and bound to the dumbbell impurity. 

Clearly, relieving the close-packing constraint employed in earlier work\cite{Gerbode2008,Gerbode2010} leads to an exciting insight, namely, that one can tune the defect induced by a dumbbell impurity by a minute change in its length. While dumbbell particles in a close-packed system can only take up three possible orientations with respect to the crystal, namely, along the three lattice directions.\cite{Gerbode2010}, the above observations indicate that this is not the case in our repulsive crystal. In the following, we will investigate the dumbbell motion in and alignment with the hexagonal crystal. 
  
\begin{figure}[t!]
	\centering
	\includegraphics[width=0.5\textwidth]{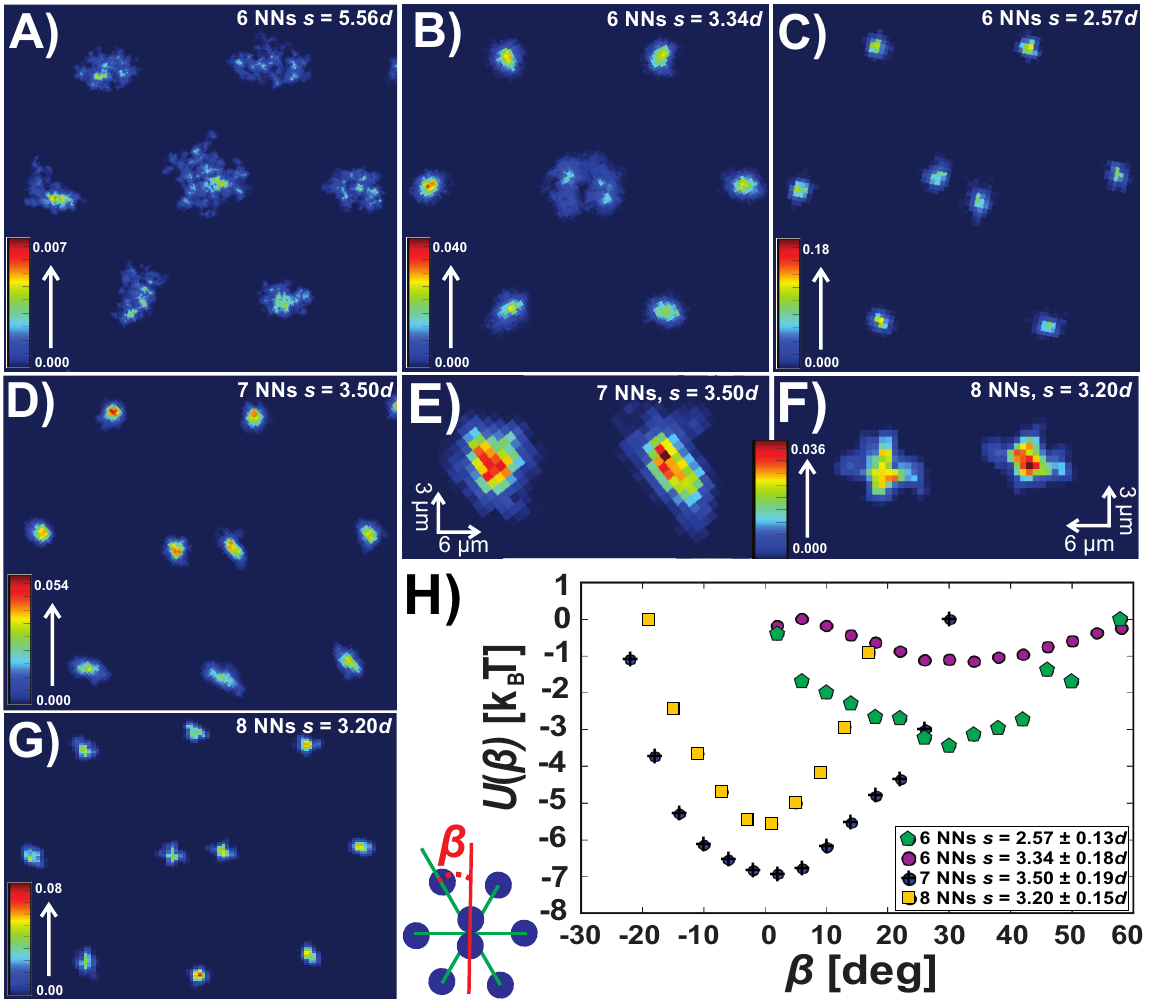}
	\caption{\footnotesize Motion and orientation of dumbbells in a hexagonal crystal of highly repulsive spheres at a fluid interface. Probability density of the position of the two spheres of dumbbells $P(r)$ surrounded by six NNs, tracked over $>$ five minutes, in a crystal with A) $s$ = (5.6 $\pm$ 0.6) $d$, B$s$ = (3.34 $\pm$ 0.18) $d$ and C) $s$ = (2.57 $\pm$ 0.13) $d$. The rotational and translational motion of dumbbell is increasingly restricted at increasing lattice spacing $s$. D) and E) $P(r)$ of a dumbbell with seven NNs at $s$ = (3.50 $\pm$ 0.19) $d$. The motion of the dumbbell is restricted and not symmetric for the two spheres. F) and G) $P(r)$ of a dumbbell with eight NNs at $s$ = (3.20 $\pm$ 0.15) $d$. The dumbbell only slightly translates in the x and y direction. H) Confinement energies of dumbbells with six, seven and eight NNs. Dumbbells with six NNs preferentially orient at $\beta$ = 30$^{\circ}$ and dumbbells with seven and eight NNs at $\beta$ = 0$^{\circ}$. Bottom left: Illustration of a hexagonal lattice with crystal axes in green. The dumbbell orientation is defined by the angle $\beta$ between the dumbbell axis and the closest axis.}
	\label{fig:ArticleLL_PositionDB}
\end{figure}

\textbf{\textit{Dumbbell dynamics}} The translational and rotational motion of dumbbell impurities is expected to depend on the lattice spacing $s$ of the surrounding crystal. With decreasing lattice spacing, the dumbbells become gradually more constrained. This implies that the area accessible to the dumbbell and the surrounding spheres decreases with decreasing $s$, see Fig. \ref{fig:ArticleLL_PositionDB}A-C and S5. The figure shows the probability density functions of the position $P(r)$ of dumbbells with six NNs over several minutes for crystals with A) $s$ = (5.6 $\pm$ 0.6) $d$, B) $s$ = (3.34 $\pm$ 0.18) $d$ and C) $s$ = (2.57 $\pm$ 0.13) $d$. At the highest $s$ the dumbbell only experiences some restrictions in the translational motion. At $s$ = (3.34 $\pm$ 0.18) $d$ the dumbbell was translationally confined to a lattice position, but full rotation around the center of mass of the dumbbell was still observed. However, the dumbbell shows preferred orientations with respect to the crystal axis. At $s$ = (2.57 $\pm$ 0.13) $d$ both the translational and rotational motion of the dumbbell were restricted by the crystal and the dumbbell aligned with its long axis in between two crystal axes. 

The translation and rotation of dumbbells with seven NNs were already limited in a crystal with $s$ = (3.50 $\pm$ 0.19) $d$, see Fig. \ref{fig:ArticleLL_PositionDB}D-E. Interestingly, the dumbbell motion is asymmetric due to the additional freedom created by the isolated dislocation. The dumbbell is preferably aligned parallel to the Burgers vector (Fig. \ref{fig:Voronoi}B2). Dumbbells with eight NNs, on the other hand, move preferentially parallel and perpendicular to the crystal orientation axis that the dumbbell long axis is aligned with (Fig. \ref{fig:ArticleLL_PositionDB}F-G). Thus, the motion pattern of the dumbbell intricately depends on its length and hence number of NNs. 
 
To further quantify the dumbbell's orientation preference, we define the angle $\beta$ as the smallest angle between its long axis and any of the crystal orientation axes, depicted in red and green, respectively, in Fig. \ref{fig:ArticleLL_PositionDB}H. We plot the confinement energy $U(\beta)$ in Fig. \ref{fig:ArticleLL_PositionDB}H by inverting the probability distribution $P(\beta)$ of datasets of single dumbbells with at least 2000 frames. 
We set $U(\beta)$=0 at the least likely value of $\beta$.
Short dumbbells with six NNs preferentially oriented at $\beta$ = 30$^{\circ}$. The energie minimum  $U(\beta)_{min}$ of the dumbbell at this angle increased from -1 $k_B$$T$ to -3.5 $k_B$$T$ when the lattice spacing decreased from $s$ = 3.34 $\pm$ 0.18$d$ to $s$ = 2.57 $\pm$ 0.13$d$. 
In contrast, dumbbells with larger $s_{DB}$ preferred a different orientation, namely $\beta=0^{\circ}$, at similar lattice spacing. 
A dumbbell with seven NNs experienced $U(\beta=0^{\circ})_{min}=-7 k_BT$. The graph has a slightly asymmetric shape, resulting from the asymmetric environment around the dumbbell. Similarly, an ensemble-based analysis of 34 dumbbells yielded an average $\beta$-value of 6.5 $\pm$ 4.7$^{\circ}$. 
Dumbbells with eight NNs explore a slightly narrower range of  -22$^{\circ}$ $<$ $\beta$ $<$ 22$^{\circ}$, with   $U(\beta=0^{\circ})_{min}= -6 k_B$$T$. An ensemble-based analysis of 22 different dumbbells yielded a similar value for  $\beta$ = 3.6 $\pm$ 2.5$^{\circ}$. Their orientation along the crystal axis agrees with earlier finding in close-packed crystals where dumbbells also feature 8 NNs \cite{Gerbode2010} and is expected for repulsive crystals.

In summary, because subtle changes in dumbbell length affect the number of NNs, they have a strong impact on the dynamic behavior of the dumbbells. 

\begin{figure*}[t!]
	\centering
	\includegraphics[width=1\textwidth]{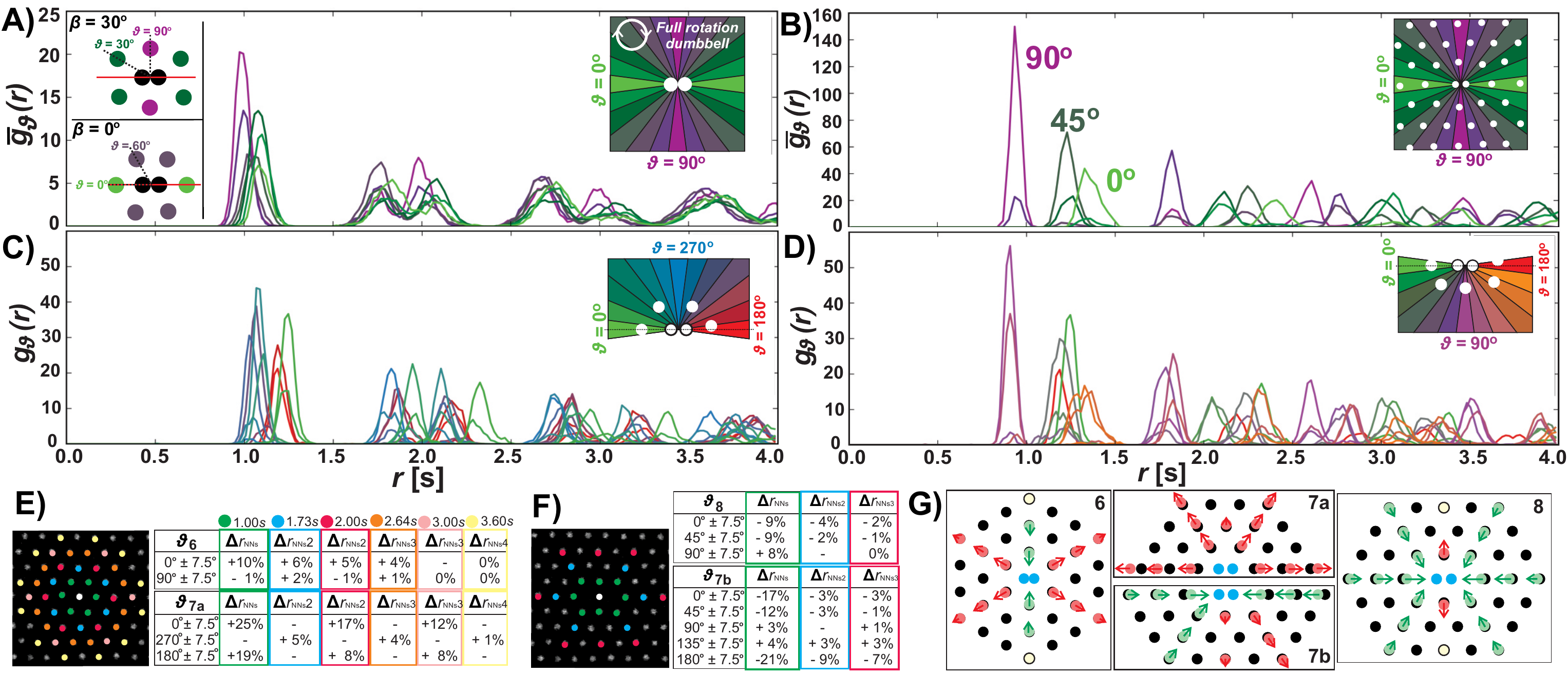}
	\caption{Crystal distortion caused by dumbbells with six ($s$ = 3.34 $d$), seven ($s$ = 3.50 $d$) and eight ($s$ = 3.20 $d$) NNs at a similar lattice spacing $s$. $\bar{g}_{\vartheta}(r)$ of spheres surrounding a dumbbell with A) six and B) eight NNs. The dumbbell with six NNs undergoes full rotation in time, while the rotational motion of the dumbbell with eight NNs is restricted. Symmetry in the crystals is maintained as peaks are observed at fixed radial distances $r$. Asymmetry is introduced by dumbbells with seven NNs resulting in long-ranged distortions. C-D) $\bar{g}_{\vartheta}(r)$ of spheres surrounding a dumbbell with seven NNs. The crystal adapts to the sphere positioning similar to the one observed for dumbbells with C) six NNs on one side of the long dumbbell axis and to the positioning observed for dumbbells with D) eight NNs on the other side of the dumbbell axis. E-F) Comparison of the position of the neighboring spheres to a perfect hexagonal crystal for dumbbells occupying E) one lattice spacing (six NNs) and F) two lattice spacings (eight NNs). The crystals are distorted only locally and order is restored at a distance of approximately $4s$. G) Illustration of the distortion of the hexagonal crystal caused by dumbbells with six, seven and eight NNs.}
	\label{fig:distortion}
\end{figure*}

\textbf{\textit{Crystal distortion}} While the crystal determines the orientation and motion of the dumbbell, the dumbbell in turn also affects the position and hence the crystal order of the surrounding spheres. We describe the position of all spheres by their distance $r=|\vec{r}|$ from the center of mass (COM) of the dumbbell and the angle $\vartheta$ between the dumbbell long axis and $\vec{r}$, (Fig. S6). To capture the anisotropic distortion, we use a radial distribution function $g_\vartheta(r)$ limited to the $ {\delta}{\vartheta}rdr$ area set by the angular range [$\vartheta$ $\pm$ $\frac{{\delta}{\vartheta}}{2}$]. 

We plot $g_\vartheta(r)$ separately for $\vartheta$ = 0$^{\circ}$, 15$^{\circ}$, 30$^{\circ}$....360$^{\circ}$ $\pm$ $\frac{{\delta}{\vartheta}}{2}$, with $\delta\vartheta$ = 15$^{\circ}$, covering the complete radial distribution (Fig. S7). We compare the dumbbells with six, seven, and eight NNs with similar lattice spacing $s$ as shown in Fig. \ref{fig:ArticleLL_PositionDB}B,D, and G. For dumbbells with six and eight NNs, we exploit the symmetry in the four quadrants observed in Fig. S7 ($\vartheta$ $\epsilon$ [0$^{\circ}$ , 90$^{\circ}$], [90$^{\circ}$, 180$^{\circ}$], [180$^{\circ}$, 270$^{\circ}$] and [270$^{\circ}$, 360$^{\circ}$]) and plot the averaged $\bar{g}_{\vartheta}(r)$ in Fig. \ref{fig:distortion}A and B, respectively. 

Because the dumbbell with six NNs can still undergo full rotation, e.g. explore the full range of $\beta$, peaks are observed at all ranges of $\theta$. We quantify the changes in the position with respect to an undistorted two-dimensional hexagonal crystal with the same lattice spacing $s$, see Fig. \ref{fig:distortion} E and F and S8. For six NN dumbbells, the NNs were positioned at $r$ = 0.99$s$ for $\vartheta$ = 90$^{\circ}$  to 1.10$s$ for $\vartheta$ = 0$^{\circ}$. These distances deviate by -1 to +10$\%$ from the expected value, $r_{hex}$ = 1$s$, for NNs in a hexagonal crystal. The full width at half maximum of the peaks $w_{DB}$ originating from the NNs is similar for all $\vartheta$-values, indicating that all NNs had similar freedom of motion. Further away from the dumbbell, the deviations from the hexagonal crystal decreased with increasing distance $r$. At $r_{hex}$ = 3.60$s$ small deviations, +1 to +2$\%$, were observed which indicates that the hexagonal order had almost completely restored. This implies that short dumbbells occupy one lattice position in the crystal and distort the hexagonal order only locally. 

In contrast, the eight NN dumbbell is much more confined (see Fig, \ref{fig:ArticleLL_PositionDB}H) and strong peaks only appear at specific $\theta$. The $\bar{g}_{\vartheta}(r)$ of dumbbells with 8 NNs exhibit a number of features stemming from the fact that these dumbbells occupy two lattice positions in a hexagonal crystal while taking up less space. The surrounding spheres clearly occupy positions expected for a hexagonal crystal, implying pronounced peaks at $\vartheta$ = 0$^{\circ}$, 45$^{\circ}$ and 90$^{\circ}$ with respect to the center of the dumbbell (Figure \ref{fig:distortion}B). However, because the point of reference is now exactly in between two normal lattice sites, the distance to the NNs in direction of the long ($\vartheta$ = 0$^{\circ}$) and short ($\vartheta$ = 90$^{\circ}$) axis of the dumbbell would be 1.5 and 0.87 lattice positions in an undisturbed hexagonal crystal, respectively. As a consequence, the NNs at $\vartheta$ = 0$^{\circ}$ are pushed by the crystal towards the dumbbell ($r=0.91\cdot r_{hex}$) and the NNs at $\vartheta$ = 90$^{\circ}$ are pushed outwards by the dumbbell ($r=1.08\cdot r_{hex}$), leading to a clear separation of the peaks in $\bar{g}_{\vartheta}(r)$. Similarly, the confinement of the surrounding spheres indicated by the full width at half maximum increased with increasing $\vartheta$, see SI. 
The deviations from a perfect hexagonal crystal quickly decreased with increasing distance $r$. Already in the third row of NNs, NNs3, differences of only -1 and -2$\%$ were measured indicating that the hexagonal order was almost completely restored. Thus, as the Voronoi diagram already predicted, see Fig. \ref{fig:Voronoi}C4, dumbbells with eight NNs occupy two lattice sites in the hexagonal crystal and distort the crystal only locally.

However, the long-ranged asymmetric distortion induced by dumbbells with seven NNs splits the crystal plane essentially into two halfs divided by the dumbbell long axis, see Fig. \ref{fig:distortion}C and D. We analysed the position of the spheres for $\vartheta$$\epsilon$[0$^{\circ}$, 180$^{\circ}$] and $\vartheta$$\epsilon$[180$^{\circ}$, 360$^{\circ}$], separately. Surprisingly, we find that on one side the spheres position as if the dumbbell was surrounded by six NNs, see Fig. \ref{fig:distortion}C. On the other side, where also the isolated dislocation is present, the spheres orient similar to the eight NNs case, see Fig. \ref{fig:distortion}D. Thus, while the dumbbell would prefer to align at $\beta=30^{\circ}$ on one side of the crystal ($\vartheta \epsilon$[0$^{\circ}$, 180$^{\circ}$]), the stronger confinement around $\beta=0^{\circ}$ imposed by the other side ($\vartheta \epsilon$[180$^{\circ}$, 360$^{\circ}$]) leads to an orientation roughly in line with the crystal axis ( $\beta=6.5\pm 4.7^{\circ}$) yet with a broader potential well, see Fig. \ref{fig:ArticleLL_PositionDB}H. As a consequence, the NNs in the range $\vartheta \epsilon$[0$^{\circ}$, 180$^{\circ}$] are pushed outwards due to the unfavourable orientation of the dumbbell with respect to the crystal. On the other half, the crystal distortion is similar to that induced by dumbbells with 8 NNs, see Fig. \ref{fig:distortion}G. Dumbbells with 7 NNs thereby behave as if they occupy an intermediate non-integer number of lattice sites between 1 and 2. We call this unprecedented behaviour occupation of 1.5 lattice sites. 

\textbf{\textit{Conclusions}}
By studying repulsive crystals with dumbbell impurities we relieved the constraints present in close-packed systems. This allowed us to continuously tune and quantify the confining effect of the surrounding crystal on the dumbbell orientation and dynamic. In turn, we discovered a striking dependence of the induced defect type and range on the length of the dumbbell, that alternates between a local and a long-range distortion of the crystal. Local distortions occur for dumbbells occupying an integer number of lattice sites, that is 1 and 2 lattice sites. In contrast, long-range distortions appeared for dumbbells that spawn dislocations by insertion of an additional semi-infinite row. These dumbbells thereby effectively occupy a non-integer number of lattice sites. We expect this behaviour will also apply to impurities with higher aspect ratios, that is, that the number of occupied lattice sites in the hexagonal crystal will alternate between integer and non-integer numbers, thereby inducing a local and a long-range distortion, respectively. This could be tested by chains of spheres that extend over at least one $s$. 

Our results emphasise that the distortions induced by anisotropic impurities differ from those stemming from isotropic impurities, which will likely also affect the physical properties of the crystal. In particular, we expect that the mechanical behavior of the crystal such as slip, fracture, and fatigue, should be affected by the particle-bound nature of the dislocation. For example, while dislocations typically weaken crystals by increasing slip, dislocation motion which underlies slip should be strongly reduced or even suppressed for particle-bound dislocations. The complex effect of doping by anisotropic impurities on the mechanical properties of the crystal will likely yield rich new behavior and exciting as-of-yet unexplored physics. 

\paragraph*{Acknowledgements} 
We thank D. ten Napel, P. Liu and A.Philipse from Utrecht University for help with the design of the interface cell, and the fine mechanical department of Leiden University and the glassblowing department of Utrecht University for building the interface cell. We also thank Roel Dullens from Oxford University for fruitful discussions.

\bibliographystyle{apsrev4-1}
\bibliography{library}

\end{document}